\def\year{2022}\relax
\documentclass[letterpaper]{article} % DO NOT CHANGE THIS
\usepackage{aaai22}  % DO NOT CHANGE THIS
\usepackage{times}  % DO NOT CHANGE THIS
\usepackage{helvet}  % DO NOT CHANGE THIS
\usepackage{courier}  % DO NOT CHANGE THIS
\usepackage[hyphens]{url}  % DO NOT CHANGE THIS
\usepackage{graphicx} % DO NOT CHANGE THIS
\usepackage{natbib}  % DO NOT CHANGE THIS AND DO NOT ADD ANY OPTIONS TO IT
\usepackage{caption} % DO NOT CHANGE THIS AND DO NOT ADD ANY OPTIONS TO IT
\DeclareCaptionStyle{ruled}{labelfont=normalfont,labelsep=colon,strut=off} % DO NOT CHANGE THIS
\usepackage{algorithm}
\usepackage{newfloat}
\usepackage{listings}

\usepackage{latexsym}

\usepackage[T1]{fontenc}
\usepackage[utf8]{inputenc}

\usepackage{microtype}

\usepackage{hyperref}
\usepackage{multirow}
\usepackage{appendix}
\usepackage{wrapfig,lipsum,booktabs}
\usepackage{color}
\usepackage{subcaption}

\usepackage{algpseudocode}% http://ctan.org/pkg/algorithmicx

\floatstyle{ruled}
\newfloat{listing}{tb}{lst}{}
\floatname{listing}{Listing}
\usepackage{amsmath,amsfonts,bm}

\def\eqref#1{equation~\ref{#1}}
\def\1{\bm{1}}

\DeclareMathAlphabet{\mathsfit}{\encodingdefault}{\sfdefault}{m}{sl}
\SetMathAlphabet{\mathsfit}{bold}{\encodingdefault}{\sfdefault}{bx}{n}

\title{Self-Supervised Audio-and-Text Pre-training with Extremely Low-Resource Parallel Data}

\author {
    Yu Kang\textsuperscript{\rm 1}, 
    Tianqiao Liu\textsuperscript{\rm 1}, 
    Hang Li\textsuperscript{\rm 1}, 
    Yang Hao\textsuperscript{\rm 1}, 
    Wenbiao Ding\textsuperscript{\rm 1, 2 \footnote{The corresponding author: Wenbiao Ding. Work done at TAL.}}
}
\affiliations {
    \textsuperscript{\rm 1} TAL Education Group, Beijing, China\\
    \textsuperscript{\rm 2} Tencent, Beijing, China\\
    \{kangyu, liutianqiao, lihang4, haoyang2\}@tal.com, darwinding@tencent.com
}

\usepackage{bibentry}
\begin{document}

\maketitle

\begin{abstract}
Multimodal pre-training for audio-and-text has recently been proved to be effective and has significantly improved the performance of many downstream speech understanding tasks. However, these state-of-the-art pre-training audio-text models work well only when provided with large amount of parallel audio-and-text data, which brings challenges on many languages that are rich in unimodal corpora but scarce of parallel cross-modal corpus. In this paper, we investigate whether it is possible to pre-train an audio-text multimodal model with extremely low-resource parallel data and extra non-parallel unimodal data. Our pre-training framework consists of the following components: (1) Intra-modal Denoising Auto-Encoding (IDAE), which is able to reconstruct input text (audio) representations from a noisy version of itself. (2) Cross-modal Denoising Auto-Encoding (CDAE), which is pre-trained to reconstruct the input text (audio), given both a noisy version of the input text (audio) and the corresponding translated noisy audio features (text embeddings). (3) Iterative Denoising Process (IDP), which iteratively translates raw audio (text) and the corresponding text embeddings (audio features) translated from previous iteration into the new less-noisy text embeddings (audio features). We adapt a dual cross-modal Transformer as our backbone model which consists of two unimodal encoders for IDAE and two cross-modal encoders for CDAE and IDP. Our method achieves comparable performance on multiple downstream speech understanding tasks compared with the model pre-trained on fully parallel data, demonstrating the great potential of the proposed method. Our code is available at: \url{https://github.com/KarlYuKang/Low-Resource-Multimodal-Pre-training}.
\end{abstract}

\vspace{-0.5cm}
\section{Introduction}
\label{sec:intro}
The recent rise of large-scale unsupervised pre-training models, i.e. BERT \cite{devlin2018bert}, RoBERTa \cite{liu2019roberta}, Wav2Vec \cite{SchneiderBCA19}, Mockingjay \cite{liu2020mockingjay}, etc., in artificial intelligence communities like natural language processing (NLP) and speech signal processing (SSP) has demonstrated the effectiveness of pretraining-finetuning framework. Subsequent study like CTAL \cite{li2021ctal} successfully extends this framework to the multimodal field through designing cross-modal self-supervised tasks on the audio-and-text parallel corpus, and has significantly improved the performance of many downstream tasks at the intersection of audio and text which we denote as speech understanding tasks in the following.

However, different from collecting non-parallel unimodal corpus, the collection of parallel corpus like LibriSpeech always inquires additional manual filtering or annotating works. Due to this fact, the scale of available parallel corpus is limited comparing to the existing unimodal corpus, which restricts the benefits of large scale pre-training process.

In this paper, in order to get rid of the dependence of audio-and-text pre-training on parallelism between text and audio data, we present a novel multimodal pre-training framework mainly based on non-parallel unimodal corpora, which incorporates three components: (1) Intra-modal Denoising Auto-Encoding (IDAE), (2) Cross-modal Denoising Auto-Encoding (CDAE) and (3) Iterative Denoising Process (IDP). IDAE \cite{ArtetxeLAC18} is an effective algorithm for self-supervised training, through restoring the corrupted inputs in a unimodal manner, the model becomes capable to extract useful intra-modality information. CDAE is an extension of IDAE, which reconstructs the input text (audio), given both a noisy version of the input text (audio) and the corresponding translated noisy audio features (text embeddings). By learning to exploit the complementary information between noisy cross-modal inputs, this reconstruction process allows the model to learn inter-modal connections efficiently.

To construct the pseudo-parallel cross-modal inputs for CDAE, we propose a novel training procedure named IDP, which is inspired by the back-translation \cite{SennrichHB16,he2016dual} in neural machine translation. Back-translation is one of the most effective ways to leverage monolingual data for translation. Unlike back-translation, which regenerates pseudo-parallel data in each training round. IDP only performs one generation at the beginning of training, and applies the current model to eliminate the noise in the generated data in each subsequent training round to get pseudo-parallel inputs with less noise. IDP actually applies the reconstruction capabilities learned during CDAE, then the new pseudo-parallel data from IDP is used to further train CDAE at the next iteration, until convergence of the algorithm.

With the help of these components, our pre-training model is capable to learn both intra-modality and inter-modality connections with large-scale non-parallel unimodal corpus. To demonstrate its effectiveness, we apply our pre-training model to three established speech understanding tasks: emotion classification \cite{busso2008iemocap}, sentiment analysis \cite{zadeh2018multimodal} and speaker verification \cite{panayotov2015librispeech}. The empirical results show that our method outperforms all baselines. The main contributions of our paper are listed as follows:
\begin{itemize}
    \vspace{-0.1cm}
    \item We present a novel multimodal pre-training method with large-scale non-parallel unimodal corpus for strong audio-and-text representations including both intra-modality and inter-modality connections. We are the first to introduce multimodal pre-training in the low-resource scenarios which scarce of parallel cross-modal data.
    % \vspace{-0.1cm}
    \item Comprehensive empirical evidence demonstrates that our pre-training model outperforms all baselines on various downstream speech understanding tasks, such as emotion classification, sentiment analysis, and speaker verification. We conduct detailed ablation studies and analysis to prove the effectiveness of our pre-training strategies.
    %We demonstrate the effectiveness of our proposed pre-training framework via three established audio-and-language tasks: emotion classification, sentiment analysis and speaker verification, and our method outperforms all baselines by a large margin. 
    %We also present detailed ablation studies to prove that all of the proposed components significantly contribute to this promising results.
    %And our model achieves state-of-the-art results on all three downstream tasks.
    % \vspace{-0.1cm}
    \item Further experiments show that our pre-training method can be effective in semi-supervised manner as well. That is, in scenarios where sufficient parallel cross-modal data is available, adding extra non-parallel unimodal data can further improve the performance.
    %Comprehensive ablation studies are conducted with each component of our method to prove their effectiveness. And 
\end{itemize}

% Then conventional cross-modal de-noise pre-training tasks are applied with original introduced single modality corpus and pseudo parallel contents. To enhance the performance of CDB, the pseudo parallel contents generated by DCT is expected to be accurate. To achieve that goal, we propose the IPU, which iteratively updates the pseudo parallel contents through DCT 

% And the conventional back-translation models solve this problem through utilizing available high-performance translation models trained with large size parallel dataset. 

% Unfortunately, the precise translations between two modalities usually inquires the large amount of 

% achieve desirable results
% back-translation is first proposed by XXX and 

% target at designing advanced pre-training tasks, like to enhance the pre-trained models' performance in various downstream tasks. Apart from 

% and delicate pre-training tasks, i.e. Mask Language , have been proposed by recent researches for adapting the data 

\vspace{-0.4cm}

\section{Related Work}
% \vspace{-0.2cm}
\label{sec:related}
\subsection{Multimodal Pre-training}
Inspired by the success of language pre-training like BERT \cite{devlin2018bert} 
%and RoBERTa \citep{liu2019roberta}
, the research community has started to pay more attention to pre-training in multimodal scenarios and has achieved remarkable results. There are quite a few attempts that have been made to pre-train models for vision-and-language tasks. 
%on Visual Question Answering \citep{antol2015vqa} and Visual Commonsense Reasoning \citep{zellers2019recognition} datasets. 
In general, these pre-training methods can be divided into two categories, according to their different encoder architectures. (a) Previous works like ViLBERT \cite{lu2019vilbert} and LXMERT \cite{tan2019lxmert}, apply two unimodal networks to encode input text and images respectively and adapt cross-modal interactions in a late fusion manner. (b) The other category of pre-training frameworks like VisualBert \cite{li2019visualbert} and Unicoder-VL \cite{li2020unicoder}, concatenate vision and language features as a unified single-stream input and utilize a universal encoder to learn joint multimodal representations. It is worth noting that, the authors of VL-BERT \cite{su2019vl} claim that their unified architecture outperforms the two-stream designs. However, the unified architecture may not be suitable for our pre-training method, since the universal encoder lacks of the ability to perform the translation between two modalities.

The architecture of our dual Transformer is similar to that of LXMERT and ViLBERT. However, the model structure on which our proposed pre-training method relies is not fixed. Any model structure with the capability of translating one modality to the other can work well with our approach.

While pre-training for vision-and-language has made some progresses in recent years, audio-and-text pre-training has also started to evolved recently. SpeechBERT \cite{chuang2019speechbert} proposes a pre-training audio-and-text model for spoken question answering task. However, the pre-training task in SpeechBERT relies on parallel cross-modal data with forced alignment information between words and audio signals, and the downstream task to which the pre-training model applies are too limited. CTAL \cite{li2021ctal} proposes a pre-training model to learn audio-and-text representations for multiple downstream speech understanding tasks. However, the self-supervised pre-training task introduced in CTAL still needs large-scale of parallel audio-and-text data, which is unavailable in many low-resource languages. So, in this paper, we design several truly self-supervised tasks for audio-and-text pre-training mainly based on non-parallel unimodal data.

% While pre-training for vision-and-language has made some progress, pre-training for audio-and-language representations is under-researched. The closest work is from \cite{haque2019audio}, which adapts audio as input and formulates a multitask learning problem by reconstructing text and acoustic features from a hidden speech embedding during pre-training. As for multimodal pre-training with nonparallel corpus, it has not been investigated yet.
%As for multimodal pre-training with nonparallel corpus has not been investigated.
\vspace{-0.2cm}
\subsection{Back-Translation}
\vspace{-0.1cm}
Back-translation \cite{SennrichHB16,he2016dual} in neural machine translation is a very effective data-augmentation scheme under the semi-supervised setting. A translation model is first trained on the available parallel data, then the model is used to produce translations from the extra monolingual corpus. 
%on both source and target languages
The pairs composed of these translations with their corresponding monolingual data are then used as additional training data for the original translation system to further boost the performance. The same approach is also applied by \cite{tjandra2017listening} to enhance the performance of automatic speech recognition (ASR) and text to speech (TTS) models simultaneously with extra unpaired audio and text data. However, it still relies on a number of audio-text pairs to do supervised training and get valid ASR and TTS models firstly.

Some similar works to ours are the unsupervised machine translation \cite{lample2017unsupervised,lample2018phrase,ArtetxeLAC18} with back-translation mechanism. The authors have carefully designed some initialization methods to obtain a rough translation model to replace the supervised training process in the semi-supervised setting. Then the denoising effect of language models and automatic generation of parallel data by iterative back-translation are leveraged round by round until convergence of the algorithm. \cite{ren2019almost,xu2020lrspeech} also proposed the unsupervised ASR and TTS models in the similar way.

% Back-translation is also widely used by unsupervised machine translation \cite{lample2017unsupervised,lample2018phrase,artetxe2017unsupervised}. In a nutshell, firstly a rough translation model is carefully initialized, then the pseudo-parallel data is generated by iterative back-translation using current translation model. \cite{ren2019almost,xu2020lrspeech} also leverage it in almost unsupervised text to speech (TTS) and automatic speech recognition (ASR), which can be regarded as unsupervised translation between audio and text.
%
%In this paper, we only leverage very low-resource parallel corpus, which is only a few percent of nonparallel data. So, our method is more of a unsupervised learning than a semi-supervised learning.

%Back-translation is also widely used by unsupervised machine translation \citep{lample2017unsupervised,lample2018phrase,artetxe2017unsupervised}. In a nutshell, first a rough translation model is carefully initialized, then generate the parallel data by iterative back-translation. \citet{ren2019almost,xu2020lrspeech} also leverage it in almost unsupervised text to speech (TTS) and automatic speech recognition (ASR), which can be regarded as unsupervised translation between audio and text.

In our work, (1) the IDP leverages the denoising capability of model learned during the training process to iteratively eliminate the noise in the translated text embeddings (audio features), instead of regenerating new translation results each round like back-translation. And (2) our cross-modal encoder is able to learn bidirectional audio-and-text representation which is important for finetuning on downstream tasks, while the autoregressive decoder in previous works is only able to learn representation with unidirectional information. (3) During back-translation, the autoregressive decoder must generate translations step by step, which is very slow for long sequences like audio signals. Therefore, this is unacceptable in large-scale pre-training scenarios. In contrast, IDP can take full advantage of the GPU to eliminate noise in parallel.

% Our proposed Iterative Denoising Process (IDP) is inspired by back-translation but differs from it in two main ways: (1) in the IDP, we generate the transformed sequences for this iteration based on both the previous version and the corresponding unimodal input. While in the back-translation, 
% %only rely on the denoising effect of language model and 
% translated data is generated from scratch in each round; (2) we carefully design the corrupt function, which is similar to that of RoBERTa, for both DAE and CDAE to enable the model to learn the ability of extracting bi-directional cross-modal representations and the denoising effect simultaneously. In contrast, the models in unsupervised machine translation are sequence-to-sequence frameworks, which are difficult to be used to pre-train audio-and-language representations for downstream speech understanding tasks.

%Our method is carefully designed for audio-and-language pre-training with low-resource parallel corpus, different from machine translation where the generation of parallel data using back-translation is straightforward with sequence-to-sequence model, leveraging the back-translation with bi-directional encoders is more challenging. 
%and we demonstrate detailed experiments that our designed components are necessary to develop the capability of cross-modal pre-training with only few paired data.

\vspace{-0.2cm}

\section{Approach}
% \vspace{-0.2cm}
\label{sec:approach}
%In this section, we first describe the architecture of our multimodal framework, and then we introduce how we pre-train it with low-resource parallel data.
In this section, we first describe the model architecture of the dual Transformer, and then we detail our pre-training strategies.
\vspace{-0.3cm}
\subsection{Model Architecture}

\begin{figure*}[!hbpt]
    % \vspace{-0.5cm}
    % \captionsetup{font=small}
    \small
    \begin{subfigure}{0.3775\linewidth}
      \centering
      \includegraphics[width=0.98\linewidth]{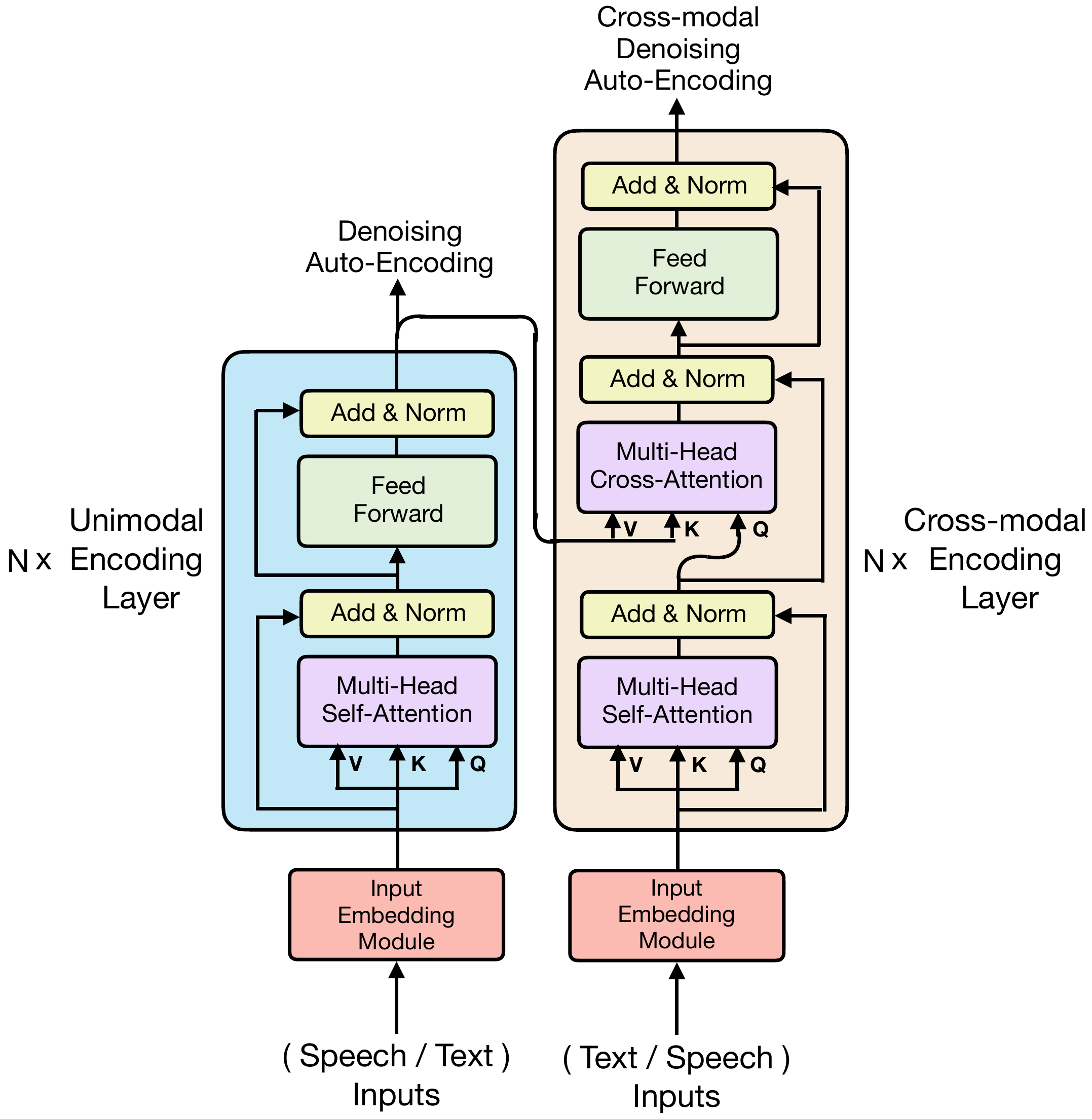}
      \caption{Model Architecture}
      \label{fig:model_architecture}
    \end{subfigure}
    \begin{subfigure}{0.6125\linewidth}
      \centering
      \includegraphics[width=0.98\linewidth]{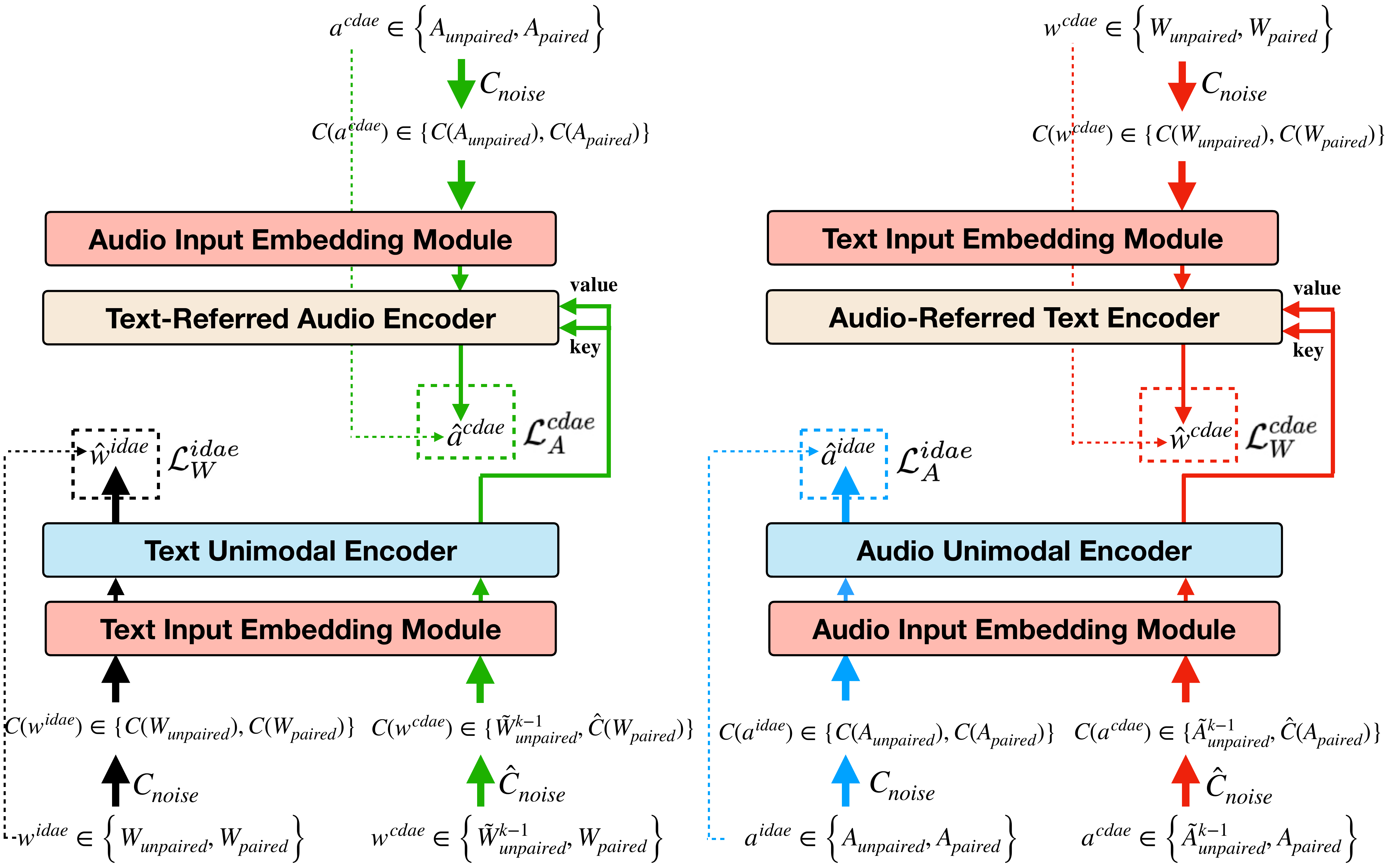}
      \caption{Pre-training Procedure}
      \label{fig:pretrain_procedure}
    \end{subfigure}
    \vspace{-0.2cm}
    \caption{(a): The structure of each encoder in our unimodal encoders and cross-modal encoders. (b): The overall pre-training flow of our method, which consists of the intra-modal denoising auto-encoding (IDAE) and the cross-modal denoising auto-encoding (CDAE), it is worth noting that, we do not present the iterative denoising process (IDP) in the figure. The solid lines denote the forward propagation and the dash lines denote the self-supervised signals.}
    \label{fig:fig1}
    \vspace{-0.4cm}
\end{figure*}

We adapt dual Transformer as our backbone model, which has been verified to be an effective cross-modal architecture by \cite{tan2019lxmert} and \cite{lu2019vilbert}, and we extend the original implementation to suit our training procedure as shown in Figure~\ref{fig:model_architecture}. It consists of three subcomponents, the input embedding module, the unimodal encoder and the cross-modal encoder, each corresponding to two different modalities, and all modules have separate parameters. We will dive into details in the following words.
%As shown in Figure~\ref{fig:model_architecture}, we build our model composed of two unimodal encoders and two cross-modal encoders, mostly on the basis of two kinds of attention layers: self-attention layers and cross-modal attention layers. 
%In this paper, we follow the formula and notations proposed by \citep{vaswani2017attention}. We adapt $\mathbf{Q}$, $\mathbf{K}$ and $\mathbf{V}$ as queries, keys and values for attention mechanism, MultiHead($\mathbf{Q}$, $\mathbf{K}$, $\mathbf{V}$) as multi-head attention, FFN($\mathbf{X}$) as position-wise feed forward networks and LayerNorm($\mathbf{X}$) as layer normalization. Next, we discuss each module in our framework in detail.
\vspace{-0.1cm}
\subsubsection{Input Embedding Module}
\label{subsec:input_module}
The input embedding module for text consists of a token embedding layer and a position embedding layer. We follow the text preprocessing of RoBERTa \cite{liu2019roberta} to encode the input text to token embeddings for text modality.

The input module for audio is composed of a position embedding layer and a dense layer which projects the audio features to hidden size. Following \cite{chi2021audio}, each input audio signal is first transformed into frames of width 50ms and step 12.5ms. Then the 80-dimension Mel-spectrograms are extracted from each frame and concatenated with their first order derivatives, making the feature dimension to 160. Finally, we feed these features to the dense layer and add them with the position embeddings to obtain the audio embeddings for audio modality.
\vspace{-0.1cm}
\subsubsection{Unimodal Encoders}
We apply the original implementation of the Transformer encoder to encode the input embeddings to the unimodal representations for text and audio separately, named Text Unimodal Encoder and Audio Unimodal Encoder.
%After the embedding modules, we first apply two original Transformer encoders \citep{vaswani2017attention}, i.e., a $\mathbf{Text\ Encoder}$ and a $\mathbf{Audio\ Encoder}$, and each of them only focuses on a single modality (i.e. text or audio). 
%As shown in Figure~\ref{fig:}, a unimodal encoding layer consists of one multi-head self-attention sublayer and one position-wise feed forward sublayer. We stack $N$ such unimodal encoding layers in each of our unimodal encoders, using the output of $(k-1)$-th layer as the input to the $k$-th layer, and we initialize $\mathbf{U}_m^0$ with the outputs of corresponding embedding module. 
%We obtain our unimodal representations for $k$-th layer with the followings:
% \begingroup\makeatletter\def\f@size{10}\check@mathfonts
% \def\maketag@@@#1{\hbox{\m@th\large\normalfont#1}}%
% \begin{align*}
% \mathbf{\hat{U}}_m^k &= MultiHead(\mathbf{U}_m^{k-1}, \mathbf{U}_m^{k-1}, \mathbf{U}_m^{k-1})
% \\
% \mathbf{\tilde{U}}_m^k &= LayerNorm(\mathbf{\hat{U}}_m^k + \mathbf{U}_m^{k-1})
% \\
% \mathbf{U}_m^k &= LayerNorm(FFN(\mathbf{\tilde{U}}_m^k)+\mathbf{\tilde{U}}_m^k)
% \\
% m &\in \left \{ text, audio \right \} 
% \end{align*}
% \endgroup
%We denote $\mathbf{U}_m^{N} \in \mathbb{R}^{\mathcal{L} \times d_m}$ the final output from our unimodal encoder, where $m$ denotes different modalities, $d_m$ denotes the hidden size of the unimodal representations and $\mathcal{L}$ is the length of output.
\vspace{-0.5cm}
\subsubsection{Cross-modal Encoders}
Then, we apply two cross-modal encoders, named Audio-Referred Text Encoder and Text-Referred Audio Encoder respectively, each encoder includes $N$ identical layers. We input the embeddings of one modality to the first layer, and by stacking $N$ these layers, we can get the final cross-modal representations (i.e. text-referred audio representations or audio-referred text representations). 
%Each layer consists of a self-attention sub-layer, a cross-modal attention sub-layer and a feed-forward sub-layer. 
Inside each layer, the self-attention sub-layer is first applied to learn the intra-modality representation, 
% \begingroup\makeatletter\def\f@size{10}\check@mathfonts
% \def\maketag@@@#1{\hbox{\m@th\large\normalfont#1}}%
% % \vspace{-0.2cm}
% \begin{align*}
% \mathbf{\hat{H}}_m^k &= MultiHead(\mathbf{H}_m^{k-1}, \mathbf{H}_m^{k-1}, \mathbf{H}_m^{k-1})
% \\
% \mathbf{\tilde{H}}_m^k &= LayerNorm(\mathbf{\hat{H}}_m^k + \mathbf{H}_m^{k-1})
% \end{align*}
% \endgroup
then we apply cross-modal attention sub-layer which accepts the final representations from the unimodal encoder of the other modality as key and value to learn the inter-modality interactions.
% \begingroup\makeatletter\def\f@size{10}\check@mathfonts
% \def\maketag@@@#1{\hbox{\m@th\large\normalfont#1}}%
% \begin{align*}
% \mathbf{\bar{H}}_m^k &= MultiHead(\mathbf{\tilde{H}}_m^k, \mathbf{U}_{\tilde{m}}^N, \mathbf{U}_{\tilde{m}}^N)
% \\
% \mathbf{\ddot{H}}_m^k &= LayerNorm(\mathbf{\bar{H}}_m^k+\mathbf{\tilde{H}}_m^k)
% \\
% \mathbf{H}_m^k &= LayerNorm(FFN(\mathbf{\ddot{H}}_m^k)+\mathbf{\ddot{H}}_m^k)
% \\
% m &\in \left \{ text, audio \right \} 
% \\
% \tilde{m} &\in \left \{ audio, text \right \} 
% \end{align*}
% \endgroup
Finally, we obtain the audio-referred text representations $\boldsymbol{H}_{w}^{N} \in \mathbb{R}^{\mathcal{T}_{w} \times d}$ and text-referred audio representations $\boldsymbol{H}_{a}^{N} \in \mathbb{R}^{\mathcal{T}_{a} \times d}$, 
%which includes both the intra-modal interactions of specific modality $m$ and the inter-modal interactions between $m$ and $\tilde{m}$, 
where $d$ denotes the hidden size of the representations. $\mathcal{T}_{w}$ and $\mathcal{T}_{a}$ are the output lengths for each modality.
\vspace{-0.2cm}
\subsection{Pre-training Problem Statement}
We consider a dataset which consists of two large-scale non-parallel unimodal corpora and a limited parallel cross-modal corpus. We denote the non-parallel unimodal corpora for text and audio as $\mathcal{W}_{unpaired}$ and $\mathcal{A}_{unpaired}$ respectively, and these two corpora do not correspond to each other. Meanwhile, the parallel corpus includes a small set of paired data $(\mathcal{W}_{paired}, \mathcal{A}_{paired})$. Then, our objective is to conduct a well performing audio-and-text pre-training model with the above dataset.
%We consider a dataset which consists of a large-scale unpaired corpus and a low-resource paired corpus. In the unpaired corpus, there are a set of sentences, denoted by $\mathcal{W}_{unpaired}$, and another set of audio, denoted by $\mathcal{A}_{unpaired}$. These two sets do not correspond to each other. Meanwhile, the paired corpus includes a set of parallel data $(\mathcal{W}_{paired}, \mathcal{A}_{paired})$. Then, we will pre-train our model using this dataset.
\vspace{-0.2cm}
\subsection{Intra-modal Denoising Auto-Encoding}
\label{subsec:idae}
Given the large amount of non-parallel audio and text data, building capabilities of learning context-sensitive unimodal representations and reconstructing corrupted inputs is of great importance. To this end, we leverage denoising auto-encoder \cite{vincent2008extracting} to reconstruct the audio and text inputs from the corrupted version of itself, which is widely used in self-supervised learning. The objective function $\mathcal{L}^{idae}$ of the intra-modal denoising auto-encoding (IDAE) is formulated as:
\vspace{-0.1cm}
%Given the large amount of unpaired audio and text data, building the capability of extracting intra-modality connections is the first step. To this end, we leverage denoising auto-encoder \citep{vincent2008extracting} to reconstruct the audio and text inputs from the corrupted version of itself, which is a typical way of self-supervised learning. The objective function $\mathcal{L}^{dae}$ of the denoising auto-encoding (DAE) is formulated as follows:
\begingroup\makeatletter\def\f@size{10}\check@mathfonts
\def\maketag@@@#1{\hbox{\m@th\large\normalfont#1}}%
% \small
\begin{align*}
\mathcal{L}^{idae} &= \mathcal{L}_{W}(w|C(w);\theta_{uni}^{W};w \in \left \{ \mathcal{W}_{unpaired}, \mathcal{W}_{paired} \right \} )
\\
&+\mathcal{L}_{A}(a|C(a);\theta_{uni}^{A};a \in \left \{ \mathcal{A}_{unpaired}, \mathcal{A}_{paired} \right \} )
\end{align*}
\endgroup

\noindent where $\theta_{uni}^{W}$ and $\theta_{uni}^{A}$ denote the model parameters of the text unimodal encoder and the audio unimodal encoder respectively. $C$ is a corrupt function, where for text modality, we dynamically select some input tokens with a probability of 15\% and the selected tokens are replaced with a special <mask> token 80\% of the time, a random token 10\%, and unaltered 10\%. For audio modality, the corrupt function is a little bit different, at first, we split the audio features in separate segments according to $S_{num}$ successive frames per segment, where $S_{num}$ is uniformly sampled from $20$ to $50$. Then we randomly select 15\% of these segments and for each of them, we mask it all to zero 80\% of the time, replace it with the other $S_{num}$ randomly selected frames within the audio 10\% of the time, and keep it unchanged for the remaining cases. In this manner, we prevent the model from exploiting local smoothness of acoustic frames and the model is required to reconstruct inputs based on global information rather than local messages. $\mathcal{L}_{W}$ and $\mathcal{L}_{A}$ denote the loss for text and audio respectively, and we implement them as:
% \vspace{-0.2cm}
%\noindent where $\theta_{uni}^{W}$ and $\theta_{uni}^{A}$ denote the model parameters of the unimodal text encoder and the audio encoder respectively. $C$ is a noise function, for text modality, we dynamically select some input tokens with a probability of 15\% and the selected tokens are replaced with a special <mask> token 80\% of the time, a random token 10\%, and unaltered 10\%, which is almost the same as the setup in RoBERTa. For audio modality, the noise function is a little bit different, at first, we split the audio in separate segments according to $C_{num}$ consecutive frames per segment, where $C_{num}$ is uniformly sampled from $20$ to $50$. Then we randomly select 15\% of these segments and for each of them, we mask it all to zero 80\% of the time, replace it with the other $C_{num}$ randomly selected frames within the audio 10\% of the time, and keep it unchanged for the remaining cases. In this manner, we prevent the model exploiting local smoothness of acoustic frames and the model is required to reconstruct inputs based on global information rather than local messages. 
%
%$\mathcal{L}_{W}$ and $\mathcal{L}_{A}$ denote the objective function for text and audio respectively, and we implement them as:
\begingroup\makeatletter\def\f@size{10}\check@mathfonts
\def\maketag@@@#1{\hbox{\m@th\large\normalfont#1}}%
% \small
\begin{align*}
\mathcal{L}_{W}(y|x;\theta_{uni}^{W}) &= - \log{P(y|x;\theta_{uni}^{W})}
\\
\mathcal{L}_{A}(y|x;\theta_{uni}^{A}) &= L1(y, f(x;\theta_{uni}^{A}))
\end{align*}
\endgroup

\noindent where $L1$ denotes the L1 loss function. It is worth noting that although the $\mathcal{L}_{W}$ and $\mathcal{L}_{A}$ have different scales of loss, our attempts to balance them do not have any effect on the final performance, so we choose to leave them unchanged for all experiments.

\vspace{-0.2cm}
\subsection{Iterative Denoising Process}
Constructing pseudo-parallel data without introducing too much noise is the key component in leveraging the large-scale non-parallel corpora for cross-modal pre-training. For this purpose, we design an iterative denoising process (IDP) to perform modality translation with step-by-step noise reduction. We can regard the dual Transformer as a text-to-audio Transformer and a audio-to-text Transformer. In order to translate text to audio, we input the unimodal text corpus to the text unimodal encoder and input the translated audio features from last iteration to the text-referred audio encoder. These two encoders make up the text-to-audio Transformer. Similarly, for audio-to-text translation, we switch to use the audio-to-text Transformer which composed of the audio unimodal encoder and the audio-referred text encoder, accepting unimodal audio corpus and translated text embeddings from last iteration as inputs respectively. Before the start of each iteration in the pre-training phase, we first translate the non-parallel unimodal text (audio) data $w$ ($a$) into less-noisy $\tilde{a}$ ($\tilde{w}$). Specifically, we obtain the translated results for the k-th training iteration with the followings:
\vspace{-0.1cm}
\begingroup\makeatletter\def\f@size{10}\check@mathfonts
\def\maketag@@@#1{\hbox{\m@th\large\normalfont#1}}%
% \small
\begin{align*}
\tilde{w}_{k} &= f(\tilde{w}_{k-1},f(a;\theta_{uni}^{A});\theta_{cross}^{W})
\\
\tilde{a}_{k} &= f(\tilde{a}_{k-1},f(w;\theta_{uni}^{W});\theta_{cross}^{A})
\end{align*}
\endgroup

\noindent where $\theta_{cross}^{W}$ and $\theta_{cross}^{A}$ denote the model parameters of the audio-referred text encoder and the text-referred audio encoder respectively. $f(;\theta)$ denotes one forward propagation of an encoder in the dual Transformer, and $\theta$ determines exactly which encoder it is.

Instead of getting raw text or raw audio signal after modality translation, $\tilde{w}_{k}$ and $\tilde{a}_{k}$ should retain information of $a$ and $w$ as much as possible. To this end, for text modality we output embeddings rather than tokens, and for audio modality we output the acoustic features mentioned in input embedding module section.
\vspace{-0.2cm}
\subsection{Warm-Up Stage}
To jump start the process, we suggest a warm-up stage at the very beginning of training to obtain a rough but reasonable initial translation results, and in the subsequent training process, we use the IDP to gradually eliminate the noise in it. 

Firstly, we train the text-to-audio Transformer and audio-to-text Transformer with the low-resource parallel data $(\mathcal{W}_{paired}, \mathcal{A}_{paired})$ as follows:
% The remaining question is, how to initialize the model with the capability of transforming one modality to the other. So, we start with a warm-up stage, by using the low-resource paired data as a guidance, to train the model with the objective of modality transforming, which can be formulated as:
\vspace{-0.1cm}
\begingroup\makeatletter\def\f@size{10}\check@mathfonts
\def\maketag@@@#1{\hbox{\m@th\large\normalfont#1}}%
% \small
\begin{align*}
\mathcal{L}^{warm} &= \mathcal{L}_{W}(w|a,w_{masked};\theta_{uni}^{A},\theta_{cross}^{W})
\\
&+ \mathcal{L}_{A}(a|w,a_{masked};\theta_{uni}^{W},\theta_{cross}^{A})
\end{align*}
\endgroup

\noindent where $(w,a) \in (\mathcal{W}_{paired}, \mathcal{A}_{paired})$, $w_{masked}$ ($a_{masked}$) is the sequence of special mask tokens (segments) with the equal length to $w$ ($a$). For training the text-to-audio Transformer, we input $w$ to the text unimodal encoder and input $a_{masked}$ to the text-referred audio encoder to reconstruct $a$. Training the audio-to-text Transformer is similar.
% $w_{masked}$ and $a_{masked}$ are sequences of the same length as $w$ and $a$, which are full of masks. 
%We input $w_{masked}$ and $a_{masked}$ to the audio-referred text encoder and the text-referred audio encoder respectively, and constrain the cross-modal encoders to reconstruct $w$ and $a$ only conditioned on the features of other modality provided by the unimodal encoders.
In this way, the model is constrained to perform modality translation based solely on the source modality. After the initial training, we initialize $\tilde{w}_{0}$ and $\tilde{a}_{0}$ for unimodal non-parallel corpora as follows:
% After warm-up training, we use the model to transform the sequences in the unimodal corpus to the other modality:
\vspace{-0.2cm}
\begingroup\makeatletter\def\f@size{10}\check@mathfonts
\def\maketag@@@#1{\hbox{\m@th\large\normalfont#1}}%
% \small
\begin{align*}
\tilde{w}_{0} &= f(w_{masked},f(a;\theta_{uni}^{A});\theta_{cross}^{W})
\\
\tilde{a}_{0} &= f(a_{masked},f(w;\theta_{uni}^{W});\theta_{cross}^{A})
\end{align*}
\endgroup

\noindent where $w \in \mathcal{W}_{unpaired}$ and $a \in \mathcal{A}_{unpaired}$. 

It's worth noting that since we do not know the exact length of the ground truth, we set the lengths of $w_{masked}$ and $a_{masked}$ as two hyper-parameters which actually limit the maximum amount of information that can be included in the translation results. And we set 256 for text and 1000 for audio according to experiments. If the exact length of the ground truth is much longer than the length hyper-parameter, the model should learn to retain as much important information as possible during the translation process, and similarly, if the exact length is much shorter than the length hyper-parameter, then the model should learn to use something like padding information to fill the information space.

\vspace{-0.2cm}
\subsection{Cross-modal Denoising Auto-Encoding}
\label{subsec:cdae}

We extend the intra-modal denoising auto-encoding (IDAE) to cross-modal denoising auto-encoding (CDAE) for learning inter-modality connections and cross-modal denoising capability. We train the model by reconstructing the input text (audio), given both a noisy version of the same input text (audio) and the noisy translated audio features (text embeddings) from IDP.
%By extending the denoising auto-encoding (DAE) to cross-modal denoising auto-encoding (CDAE), we constrain the model to be able to reconstruct inputs of one modality conditioned on the noisy version of both modalities.
\vspace{-0.1cm}
\subsubsection{Non-parallel Corpus}
For training the text-to-audio Transformer, we input $\tilde{w}$ to the text unimodal encoder and input corrupted $a$ to the text-referred audio encoder to reconstruct $a$. As for training the audio-to-text Transformer, we switch to use $(\tilde{a}, w)$ as the pseudo-parallel inputs. The objective function is formulated as follows:
% Inter-Modality translation is the key component in leveraging the large-scale unpaired corpus for cross-modal pre-training. In this phase, we adopt our framework as a bidirectional translation model between two modalities. In order to translate text to audio, we input the unimodal text corpus to the text unimodal encoder and input the translated audio features at last iteration to the text-referred audio encoder. We can regard these two encoders as a text-to-audio module. For audio-to-text translation, we switch to use the audio unimodal encoder and the audio-referred text encoder, which accepts unimodal audio corpus and translated text embeddings. Before the start of each iteration in the pre-training phase, we first translate the unpaired unimodal text (audio) data $w$ ($a$) into $\tilde{a}$ ($\tilde{w}$), then we train the text-to-audio (audio-to-text) module on the transformed pair $(\tilde{w}, a)$ ($(w, \tilde{a})$) to perform CDAE pre-training, which is inspired by back-translation \cite{lample2017unsupervised} in neural machine translation. The objective function is formulated as follows:
% \vspace{-0.3cm}
%\noindent \textbf{Unpaired Corpus} \indent Modality transforming is the key component in leveraging the large-scale unpaired corpus to learn inter-modality connections, the proposed cross-modal encoders have the ability to do this well, and the details will be described in following section. The principle here is to regard the unimodal data and its corresponding modality-transformed sequences as parallel data, the objective function is formulated as follows:
\begingroup\makeatletter\def\f@size{10}\check@mathfonts
\def\maketag@@@#1{\hbox{\m@th\large\normalfont#1}}%
% \small
\begin{align*}
\mathcal{L}^{cdae}_{unpaired} &= \mathcal{L}_{W}(w|\tilde{a},C(w);\theta_{uni}^{A},\theta_{cross}^{W})
\\
&+ \mathcal{L}_{A}(a|\tilde{w},C(a);\theta_{uni}^{W},\theta_{cross}^{A})
\end{align*}
\endgroup

\noindent We adapt almost the same corrupt function $C$ as described in IDAE section. However, in order to prevent the model from reconstructing inputs based only on itself and ignoring the cross-modal information, we increase the probability of corruption in $C$ from 15\% to 30\%.

The intuition behind CDAE process is that as long as the initial translation results $\tilde{w}_{0}$ and $\tilde{a}_{0}$ retain at least some information of the source modalities, the unimodal encoders will map such translation into a representation in feature space that also corresponds to a cleaner version of the input, since the unimodal encoders are trained to denoise in IDAE. At the same time, the cross-modal encoders are trained to predict noiseless outputs, conditioned on noisy cross-modal inputs in CDAE. Putting these two pieces together will eliminate noise in the translations, which will enable better IDP outputs at the next iteration, and so on so forth.

%\noindent where $\tilde{a}$ is transformed from $w \in \mathcal{W}_{unpaired}$ and $\tilde{w}$ is transformed from $a \in \mathcal{A}_{unpaired}$. $\theta_{cross}^{W}$ and $\theta_{cross}^{A}$ denote the model parameters of the audio-referred text encoder and the text-referred audio encoder respectively. In here, $C$ is the same type of noise function as that of Section~\ref{subsec:dae}. In order to prevent the model from reconstructing inputs based only on itself and ignoring the cross-modal information, we increase the selecting probability in $C$ from 15\% to 30\%. Furthermore, since the types of noise during reconstructions are different between the CDAE and iterative denoising process (IDP), which will be discussed in detail in the following, in order to alleviate this gap, in these 30\% selected tokens and segments, we only mask 60\% of them, replace with random tokens and segments 20\% of the time, and unaltered 20\% of the time.
\vspace{-0.1cm}
\subsubsection{Parallel Corpus}
We treat the cross-modal parallelism provided by parallel corpus as prior knowledge and use it to guide the non-parallel training process. The CDAE for parallel corpus is straightforward and the objective function is defined as follows:
\vspace{-0.1cm}
%\noindent \textbf{Paired Corpus} \indent Everything is straightforward with paired corpus:
\begingroup\makeatletter\def\f@size{10}\check@mathfonts
\def\maketag@@@#1{\hbox{\m@th\large\normalfont#1}}%
% \small
\begin{align*}
\mathcal{L}^{cdae}_{paired} &= \mathcal{L}_{W}(w|a,C(w);\theta_{uni}^{A},\theta_{cross}^{W})
\\
&+ \mathcal{L}_{A}(a|w,C(a);\theta_{uni}^{W},\theta_{cross}^{A})
\end{align*}
\endgroup

\noindent where $(w,a) \in (\mathcal{W}_{paired}, \mathcal{A}_{paired})$, and the noise function $C$ is exactly the same as that of non-parallel data. However, during non-parallel training, the input of unimodal encoder $\tilde{w}$ and $\tilde{a}$ are noisy, which shows inconsistency from the parallel corpus scenario, as the input $w$ and $a$ are noiseless. Thus we apply a new corrupt function $\hat{C}$ on input $a$ and $w$ in $\mathcal{L}_{W}$ and $\mathcal{L}_{A}$ respectively, to imitate the noise from modality translation. In detail, we first apply IDP to parallel corpus following the same procedures aforementioned, which translates $w$ ($a$) into $\tilde{a}$ ($\tilde{w}$). Then we select 30\% tokens (segments) in $w$ ($a$), and replace them with the tokens (segments) of $\tilde{w}$ ($\tilde{a}$) in the same position. 
% 80\% of the time, with random tokens (segments) 10\% of the time, and unchanged 10\% of the time.
%\noindent where $(w,a) \in (\mathcal{W}_{paired}, \mathcal{A}_{paired})$, and the noise function $C$ is totally the same as that of unpaired data. 
%
%However, since the role of low-resource paired corpus is a guidance for the unpaired corpus, it is important to constrain their training procedures in the same setting. The inputs of unimodal encoders $\tilde{w}$ and $\tilde{a}$ are noisy in the unpaired scenario, so we design a new noise function $\hat{C}$ to apply the same type of noise to the inputs of paired corpus. First, we apply modality transforming on paired corpus following the unpaired data, then, we select 30\% tokens and segments in the inputs of unimodal encoders, and replace them with the same place in the transformed sequences 80\% of the time, with random tokens and segments 10\% of the time, and unchanged 10\% of the time.
%
We re-formulate the CDAE for paired corpus as follows:
%We re-formulate the CDAE for paired corpus as follows:
\vspace{-0.2cm}
\begingroup\makeatletter\def\f@size{10}\check@mathfonts
\def\maketag@@@#1{\hbox{\m@th\large\normalfont#1}}%
% \small
\begin{align*}
\mathcal{L}^{cdae}_{paired} &= \mathcal{L}_{W}(w|\hat{C}(a),C(w);\theta_{uni}^{A},\theta_{cross}^{W})
\\
&+ \mathcal{L}_{A}(a|\hat{C}(w),C(a);\theta_{uni}^{W},\theta_{cross}^{A})
\end{align*}
\endgroup

\noindent At last, we denote $\mathcal{L}^{cdae} = \mathcal{L}^{cdae}_{unpaired} + \mathcal{L}^{cdae}_{paired}$.
%\noindent At last, we denote $\mathcal{L}^{cdae} = \mathcal{L}^{cdae}_{unpaired} + \mathcal{L}^{cdae}_{paired}$.

\vspace{-0.2cm}
\subsection{Final Objective Function}
The general procedure of our pre-training method is shown in Figure~\ref{fig:pretrain_procedure}, and the final objective function of our learning algorithm is as follows:
\vspace{-0.1cm}
\begin{align*}
\mathcal{L}(\theta_{uni}^{W},\theta_{uni}^{A}&,\theta_{cross}^{W},\theta_{cross}^{A}) = \mathcal{L}^{idae}(\theta_{uni}^{W},\theta_{uni}^{A}) 
\\
&+ \mathcal{L}^{cdae}(\theta_{uni}^{W},\theta_{uni}^{A},\theta_{cross}^{W},\theta_{cross}^{A})
\end{align*}

\vspace{-0.2cm}
\subsection{Overall Training Flow}
% The final objective function of our learning algorithm is as follows:
% \vspace{-0.2cm}
% \begingroup\makeatletter\def\f@size{10}\check@mathfonts
% \def\maketag@@@#1{\hbox{\m@th\large\normalfont#1}}%
% % \small
% \begin{align*}
% \mathcal{L}(\theta_{uni}^{W},\theta_{uni}^{A}&,\theta_{cross}^{W},\theta_{cross}^{A}) = \mathcal{L}^{dae}(\theta_{uni}^{W},\theta_{uni}^{A}) 
% \\
% &+ \mathcal{L}^{cdae}(\theta_{uni}^{W},\theta_{uni}^{A},\theta_{cross}^{W},\theta_{cross}^{A})
% \end{align*}
% \endgroup

%As explained previously, our model relies on an iterative denoising process (IDP) which starts from a warm-up stage and continuously reduces noise in the previous transformed sequences during training.

%The intuition behind our algorithm is that as long as the initial transformed sequences $\tilde{w}_{0}$ and $\tilde{a}_{0}$ include at least some information of the other modality, which is guaranteed by warm-up training, in the CDAE, the unimodal encoders will map these transformed sequences into representations of a cleaner version, since these unimodal encoders are trained to denoise the inputs in DAE. At the same time, the objective function of CDAE leads the cross-modal encoders to predict noiseless version of the inputs conditioned on both noisy inputs and the transformed sequences. Putting these two components together, in IDP, the noise in $\tilde{w}_{0}$ and $\tilde{a}_{0}$ will be gradually removed in each iteration. And it will also be beneficial to the CDAE training in the next round, and so on so forth.

The overall algorithm is described in Algorithm~\ref{alg:alg1}. A potential direction of improvement for our algorithm is that there is discrepancy of the type of noise in the inputs of cross-modal encoders between the CDAE and IDP. Specifically, the inputs of cross-modal encoders contain <mask> token for text and masked all-zero segment for audio in CDAE phase, while these special mask tokens and segments do not appear in IDP. To mitigate this discrepancy, we decrease the probability of replacing the selected time-steps with masked ones from 80\% to 60\% during CDAE pre-trainig, as a lower probability will cause the training to become unstable or even difficult to converge according to our experiment results. 
% Furthermore, since the types of noise during reconstructions are different between the CDAE and iterative denoising process (IDP), which will be discussed in detail in the following, in order to alleviate this gap, in these 30\% selected tokens and segments, we only mask 60\% of them, replace with random tokens and segments 20\% of the time, and unaltered 20\% of the time.

% As mentioned in Section~\ref{subsec:cdae}, a potential direction of improvement for IDP is that there is a mismatch of noise types between CDAE and IDP: in CDAE, the inputs of cross-modal encoders contain <mask> token for text and masked all-zero segment for audio, while in IDP, they do not appear. To mitigate this, in CDAE, we decrease the probability of replacing the selected time-steps with masked ones to 60\% according to experiments, and a lower probability will cause the training to become unstable or even difficult to converge. We have not yet figured out a complete solution, but we still found IDP helpful despite such mismatch.
\vspace{-0.2cm}
\begin{algorithm}
\small
\caption{Self-supervised Multimodal Pre-training with Low-Resource Parallel Data} 
\label{alg:alg1} 
\renewcommand{\algorithmicrequire}{\textbf{Input:}}
\renewcommand{\algorithmicensure}{\textbf{Output:}}
\begin{algorithmic}[1]

\Require 
$\left \{ \mathcal{W}_{unpaired},\mathcal{A}_{unpaired} \right \} $: large-scale non-parallel corpora; 
$(\mathcal{W}_{paired}, \mathcal{A}_{paired})$: low-resource parallel corpus; 
$K$: the total number of training epochs; 
$T$: the total number of warm-up epochs
\Ensure 
$\theta_{all}$: 
$\theta_{uni}^{W}$; 
$\theta_{uni}^{A}$;
$\theta_{cross}^{W}$;
$\theta_{cross}^{A}$

\hspace*{-36pt} \textbf{Warm Up}
\For{$t=1,T$}
    \State $\theta_{all} \leftarrow arg\min{\mathcal{L}^{warm}}$
\EndFor

\State $\tilde{w}_{0} = f(w_{masked},f(a;\theta_{uni}^{A});\theta_{cross}^{W})$
\State $\tilde{a}_{0} = f(a_{masked},f(w;\theta_{uni}^{W});\theta_{cross}^{A})$

\hspace*{-36pt} \textbf{Iterative Denoising}
\For{$k=1,K$}
    \State $\theta_{uni}^{W},\theta_{uni}^{A} \leftarrow arg\min{\mathcal{L}^{idae}}$
    \State $\theta_{all} \leftarrow arg\min{\mathcal{L}^{cdae}}$
    \State \hspace*{-2pt} \textbf{with no gradient:}
    \State \hspace*{\algorithmicindent} $\tilde{w}_{k} = f(\tilde{w}_{k-1},f(a;\theta_{uni}^{A});\theta_{cross}^{W})$
    \State \hspace*{\algorithmicindent} $\tilde{a}_{k} = f(\tilde{a}_{k-1},f(w;\theta_{uni}^{W});\theta_{cross}^{A})$
\EndFor

\State \textbf{return} 
$\theta_{all}$: 
$\theta_{uni}^{W}$; 
$\theta_{uni}^{A}$;
$\theta_{cross}^{W}$;
$\theta_{cross}^{A}$

\end{algorithmic}
\end{algorithm}

\vspace{-0.5cm}

\section{Experimental Setup and Result}
% \vspace{-0.2cm}
\label{sec:experiment}
\subsection{Pre-training Details}
\label{subsec:pretrain}

We pre-train our model on the LibriSpeech \cite{panayotov2015librispeech} dataset, which includes both audio recordings and corresponding authorized transcripts of English reading speech. We consider three subsets of LibriSpeech for pre-training: \textit{train-clean-100}, \textit{train-clean-360}, \textit{train-other-500}. In our experiments, we sample the low-resource parallel corpus from \textit{train-clean-100} subset, and in order to build the non-parallel corpus, we first combine the remaining two subsets, then split it in half and take the text from one half and the audio from the other half. In the end, we form a parallel corpus including 1 hour of speech and 500 utterances, and a non-parallel corpus including 430 hours of speech and 126k utterances.

%Following \citet{radford2019language}, we consider training a BBPE tokenizer on the LibriSpeech corpus with additional special tokens (<s>, </s>, <mask>, <pad>) as our language stream tokenizer, and we tokenize the input text into token sequence as described in Section~\ref{subsec:input_module}. For audio stream, we use Librosa \citep{mcfee2015librosa}, which is a well-established audio analysis Python package, to extract the 160-dimension input acoustic feature for each frame as described in Section~\ref{subsec:input_module}. 
For the dual Transformer, each encoder in both unimodal encoders and cross-modal encoders has 3 layers, the number of multi-head attention heads is 12 and the hidden size is 768. The total number of parameters of the whole dual Transformer is 110M.

We take Adam \cite{kingma2014adam} as our optimizer with initial learning rate of 2e-5 and a linear-decayed learning rate schedule with warm-up \cite{devlin2018bert}. We pre-train our model using 4 32G-V100 GPUs with a batch size of 8 for 500,000 steps, and the whole pre-training process takes roughly 72 hours.
\vspace{-0.2cm}
\subsection{Fine-tuning on Downstream Tasks}
We apply our pre-training model to three different kinds of widely-studied speech understanding tasks, only with simple modifications. During fine-tuning, we directly input the parallel data in downstream dataset to our model and obtain the audio-referred text representations $\boldsymbol{H}_{w}^{N} \in \mathbb{R}^{\mathcal{T}_{w} \times d}$ and text-referred audio representations $\boldsymbol{H}_{a}^{N} \in \mathbb{R}^{\mathcal{T}_{a} \times d}$ from the cross-modal encoders. Our final representation is $h^{fuse} = [\mathcal{MP}(\boldsymbol{H}_{w}^{N});\mathcal{MP}(\boldsymbol{H}_{a}^{N})] \in \mathbb{R}^{2 \times d}$, where $\mathcal{MP}$ is mean pooling over the output length.
%We apply the simple but effective mean pooling to both cross-modal representations then concatenate them to get our final representations $h^{fuse} \in \mathbb{R}^{2 \times d}$.

%We use a batch size of 4 and fine-tune for 20 epochs over each fold with 1 32G-V100 GPU. We take AdamW \citep{loshchilov2018fixing} as the optimizer in fine-tuning stage, the learning rate is initialized as 1e-5 and we apply a cosine annealing learning rate schedule \citep{loshchilov2016sgdr} to reach the optimum.
\vspace{-0.1cm}
\subsubsection{Emotion Classification}
In emotion classification task, given a speech clip, the model is asked to predict which emotion category the speech belongings to. Here, we use the widely-used dataset IEMOCAP \cite{busso2008iemocap}. IEMOCAP is recorded across 5 sessions with 5 pairs of speakers. 
%Each dialog between two speakers contains audio, transcriptions, video, and motion-capture recordings, we only use audio and transcriptions in our study. The recorded dialogues have been sliced into utterances and labelled in 10 categories (angry, happy, sad, neutral, frustrated, excited, fearful, surprised, disgusted, other) by three annotators and utterances without any text content are filtered out in our experiment. 
We follow the settings with \cite{xu2019learning} for consistent comparisons with previous works, which 
%use four emotions (angry, happy, neutral and sad) for classification and 
perform 5-fold cross-validation over sessions. 
%where each session is used as the test set in turn and remaining as training dataset. 
We adopt the same two metrics as previous works for evaluation: weighted accuracy (WA) that is the overall classification accuracy and unweighted accuracy (UA) that is the average recall over all classes. The reported WA and UA are averaged over the 5-fold cross-validation experiments, and higher WA and UA results represent better model performances.

\begin{table}[t]
    % \vspace{-0.5cm}
    % \captionsetup{font=small}
    \footnotesize
    \centering
    % \begin{center}
    \scalebox{0.7}{
    \begin{tabular}{@{}l|c|c@{}}
        \toprule
        Methods & WA $\uparrow$ & UA $\uparrow$ \\ \midrule
        LSTM\_alignment \cite{xu2019learning} & 0.6900 & 0.7014 \\
        MRDE \cite{yoon2018multimodal} & 0.6702 & 0.6764 \\
        MHA \cite{yoon2019speech} & 0.6780 & 0.6880 \\ \midrule
        Our Method & \textbf{0.7254} & \textbf{0.7339} \\ \bottomrule
    \end{tabular}
    }
    % \end{center}
    \vspace{-0.2cm}
    \caption{Comparison to the SOTA methods on the IEMOCAP dataset.}
    \label{tab:exp_result:emotion classification}
    \vspace{-0.4cm}
\end{table}

To fine-tune on IEMOCAP, the only new parameters are weights of classification layer $\mathbf{W} \in \mathbb{R}^{4 \times (2 \times d)}$, which is applied to $h^{fuse}$. The training is driven by the cross-entropy loss between the predicted class and the gold label.

We select multiple models that claim to achieve SOTA results on IEMOCAP dataset as our baselines. 
%and to be noticed, these methods are specifically designed for this task and without pre-training. 
% more details about baselines are in Appendix.
Table~\ref{tab:exp_result:emotion classification} presents our experimental results on IEMOCAP dataset. Since some prior works experiment with different train/test split, we reimplement baseline models with their published codes and unify the split as aforementioned. Our proposed method outperforms all three baselines, obtaining 3.54\% and 3.25\% absolute improvement on WA and UA respectively over the prior state of the art.
\vspace{-0.1cm}
\subsubsection{Sentiment Analysis}
We adopt CMU-MOSEI \cite{zadeh2018multi} dataset to evaluate the sentiment analysis task, which aims to predict the degree of positive and negative sentiment. 
%MOSEI contains 23,454 movie review video clips taken from YouTube. We still use only audio and corresponding transcriptions as input in our experiments. 
%Each sample in the dataset is labeled with a sentiment score from -3 (strongly negative) to 3 (strongly positive) by human annotators. 
We follow the same experimental protocol as MuIT \cite{tsai2019multimodal}, with the same train/test data split. Unlike the emotion classification task, sentiment analysis is a regression task, so we adopt two widely-used regression metrics for evaluation: mean absolute error (MAE), and the Pearson correlation coefficient (Corr) between model's predictions and human annotations. Since the prior top results reported on the CMU-MOSEI dataset are all achieved using all three modalities, so does MulT. So we reimplement it, prune the vision-related components, and re-train the model using only audio and text information.

\begin{table}[t]
    % \vspace{-0.1cm}
    % \captionsetup{font=small}
    \footnotesize
    \centering
    % \vspace{-0.2cm}
    \scalebox{0.7}{
    \begin{tabular}{@{}l|c|c@{}}
        \toprule
        Methods & MAE $\downarrow$ & Corr $\uparrow$ \\ \midrule
        MulT \cite{tsai2019multimodal} & 0.6367 & 0.6292 \\ \midrule
        Our Method & \textbf{0.6164} & \textbf{0.6804} \\ \bottomrule
    \end{tabular}
    }
    \vspace{-0.2cm}
    \caption{Comparison to the SOTA methods on the CMU-MOSEI dataset.}
    \label{tab:exp_result_sentiment}
    \vspace{-0.2cm}
\end{table}

During fine-tuning on MOSEI, we introduce additional parameters $\mathbf{W} \in \mathbb{R}^{1 \times (2 \times d)}$ to project the final representations $h^{fuse}$ to the sentiment score, and the model is trained to minimize the L1 loss between the predicted scores and the labels. As show in Table~\ref{tab:exp_result_sentiment}, we observe that our method improves the MAE and Corr by 2.03\% and 5.12\% over MulT.
\vspace{-0.1cm}
\subsubsection{Speaker Verification}
The goal of the speaker verification task is to verify the speaker identity of an utterance through comparing it with the pre-recoded voiceprints. In the text-independent speaker verification task, text can provide much more information than just text content, as a strong audio-and-text representation will include fine-grained cross-modal information such as the way and speed of pronunciation of each word, which also contains strong speaker features. In this experiment, we adopt LibriSpeech \cite{panayotov2015librispeech} for evaluation, which consists of 7 subsets in total.
%: train-clean-100, train-clean-360, train-other-500, dev-clean, dev-other, test-clean, test-other, collected from more than 2,438 speakers. 
Following the same experiment settings with prior works \cite{wan2018generalized,jung2019rawnet}, we fine-tune our pre-trained model with all training subsets, and evaluate it with \textit{test-clean} part, which contains 40 brand new speakers to the training part. Please note here, although the train set for our speaker verification task is identical with the one we used for pre-training, the speaker identity information and \textit{test-clean} data are not released during the pre-training process. Thus, it is fair to perform comparisons between our model with other prior works. We add a two-layer dense layer and a classifier over the head of final representations $h^{fuse}$ and adopt cross-entropy loss to fine-tune it. The output size of the classifier is the same as the number of unique speakers in train set.

For evaluation, we utilize the representations before classifier as the speaker embedding of input speech recording. And the cosine distance of each paired speaker embeddings are used as the indicator for the final decision. Similar to prior studies, we report the Equal Error Rate (EER) as the evaluation metric, and lower EER represents better model performance. We choose two SOTA models as our baselines \cite{wan2018generalized, jung2019rawnet}. The comparison results are shown in Table.~\ref{tab:exp_result_speaker}. From the table, we observe that the proposed method outperforms GE2E and RawNet by 1.69\% and 0.43\% respectively.

% \vspace{-0.3cm}
\begin{table}[t]
    % \vspace{-0.5cm}
    % \captionsetup{font=small}
    \footnotesize
    \centering
    \scalebox{0.7}{
    \begin{tabular}{@{}l|c@{}}
        \toprule
        Methods & EER $\downarrow$ \\ \midrule
        GE2E \cite{wan2018generalized} & 0.0379 \\
        RawNet \cite{jung2019rawnet} & 0.0253 \\ \midrule
        Our Method & \textbf{0.0210} \\ \bottomrule
    \end{tabular}
    }
    \vspace{-0.2cm}
    \caption{Comparison to the SOTA methods on the LibriSpeech dataset.}
    \label{tab:exp_result_speaker}
    \vspace{-0.5cm}
\end{table}

\vspace{-0.3cm}

\section{Analysis}
\label{sec:analysis}
\vspace{-0.1cm}
\subsection{Ablation Studies}

\begin{table*}[!hbpt]
    % \vspace{-0.5cm}
    % \captionsetup{font=small}
    \tiny
    \centering
    \setlength{\tabcolsep}{1.4mm}{
        \begin{tabular}{@{}cccccccc|cc|cc|c@{}}
            \toprule
            \multirow{2}{*}{Settings} & \multirow{2}{*}{IDAE} & \multirow{2}{*}{CDAE} & \multirow{2}{*}{IDP} & \multirow{2}{*}{Text Outputs} & \multirow{2}{*}{Audio Outputs} & \multirow{2}{*}{\begin{tabular}[c]{@{}c@{}}Paired Data\\ (hours)\end{tabular}} & \multirow{2}{*}{\begin{tabular}[c]{@{}c@{}}Unpaired Data\\ (hours)\end{tabular}} & \multicolumn{2}{c|}{Emotion Classification} & \multicolumn{2}{c|}{Sentiment Analysis} & Speaker Verification \\ \cmidrule(l){9-13} 
             &  &  &  &  &  &  &  & WA $\uparrow$ & UA $\uparrow$ & MAE $\downarrow$ & Corr $\uparrow$ & EER $\downarrow$ \\ \midrule
            w/o Pre-training &  &  &  & $\surd$ & $\surd$ & 0 & 0 & 0.7083 & 0.7119 & 0.6586 & 0.6238 & 0.0354 \\
            w/o IDAE &  & $\surd$ & $\surd$ & $\surd$ & $\surd$ & 1 & 430 & 0.7112 & 0.7201 & 0.6405 & 0.6506 & 0.0296 \\
            w/o IDP & $\surd$ & $\surd$ &  & $\surd$ & $\surd$ & 1 & 430 & 0.7241 & 0.7297 & 0.6199 & 0.6772 & 0.0232 \\
            w/o Audio Outputs & $\surd$ & $\surd$ & $\surd$ & $\surd$ &  & 1 & 430 & 0.6998 & 0.7079 & 0.6353 & 0.6333 & \_ \\
            w/o Text Outputs & $\surd$ & $\surd$ & $\surd$ &  & $\surd$ & 1 & 430 & 0.7174 & 0.7217 & 0.6211 & 0.6655 & 0.0273 \\ \midrule
            w/o Paired Data & $\surd$ & $\surd$ & $\surd$ & $\surd$ & $\surd$ & 0 & 430 & 0.7108 & 0.7220 & 0.6424 & 0.6342 & 0.0307 \\
            w/ Less Unpaired Data & $\surd$ & $\surd$ & $\surd$ & $\surd$ & $\surd$ & 1 & 180 & 0.7129 & 0.7154 & 0.6469 & 0.6318 & 0.0335 \\ \midrule
            RoBERTa & $\surd$ &  &  & $\surd$ &  & 0 & 430 & 0.6858 & 0.7003 & 0.6560 & 0.6152 & \_ \\
            Mockingjay & $\surd$ &  &  &  & $\surd$ & 0 & 430 & 0.6701 & 0.6823 & 0.6676 & 0.6011 & 0.0320 \\
            Late Fusion & $\surd$ &  &  & $\surd$ & $\surd$ & 0 & 430 & 0.7103 & 0.7257 & 0.6409 & 0.6434 & 0.0313 \\ \midrule
            Our Method & $\surd$ & $\surd$ & $\surd$ & $\surd$ & $\surd$ & 1 & 430 & 0.7254 & 0.7339 & 0.6164 & \textbf{0.6804} & 0.0210 \\ \midrule
            w/ Fully Paired Data & $\surd$ & $\surd$ &  & $\surd$ & $\surd$ & 431 & 0 & \textbf{0.7336} & \textbf{0.7406} & \textbf{0.6045} & 0.6794 & \textbf{0.0181} \\ \bottomrule
        \end{tabular}
    }
    \vspace{-0.2cm}
    \caption{The results for performing ablation studies with proposed method. The EER is not reported for settings "w/o Audio Outputs" and "RoBERTa" because it does not make sense to perform speaker verification with only text features.}
    \vspace{-0.3cm}
    \label{tab:ablation}
    \vspace{-0.2cm}
\end{table*}
\vspace{-0.1cm}
In order to study the effectiveness of different key components in our method, we present the ablation results in Table.~\ref{tab:ablation}.
% \vspace{-0.1cm}
\subsubsection{Effect of Key Components in Pre-training Strategies and Model}
Overall, our pre-training method improves the performance across all three downstream tasks (by comparing settings "w/o Pre-training" and "Our Method"). By comparing settings "w/o IDAE" with "Our Method", we see the benefits of IDAE's denoising capability and the importance of intra-modal connections. Setting "w/o IDP" removes our proposed iterative denoising process (IDP), and in each training iteration, we regenerate the translation results only conditioned on the data of source modality, like back-translation. By comparing it with "Our Method", we observe that the model's performances decrease on all three tasks, which proves the effectiveness of IDP. Each setting of "w/o Audio Outputs" and "w/o Text Outputs" only uses the output representations from either audio-referred text encoder or text-referred audio encoder for fine-tuning, and through comparing them to "Our Method", we find each of the outputs contributes to the downstream speech understanding tasks.
\vspace{-0.1cm}
\subsubsection{Effect of Pre-training Data Size}
Setting "w/o Paired Data" removes the parallel corpus and pre-trains the model only using non-parallel corpus, by comparing it with "Our Method", we observe that the model's performances drop on all downstream tasks, which proves the importance of parallel data and its guidance to the pre-training procedure. Besides, with the increment in the size of non-parallel data, our model achieves better performances on all evaluation metrics (by comparing setting "w/ Less Unpaired Data" and "Our Method"). At last, setting "w/ Fully Paired Data" is the model pre-trained with fully parallel data (431 hours in total). We use the text data in the non-parallel corpus with their corresponding audio and combine with the parallel corpus to pre-train this model. From the results, we find "Our Method" can achieve comparable performance across all three tasks and even better on correlation coefficient in sentiment analysis.
\vspace{-0.1cm}
\subsubsection{Comparisons with Unimodal Pre-training Models}
Then, setting "RoBERTa" \cite{liu2019roberta} and setting "Mockingjay" \cite{liu2020mockingjay} are unimodal pre-training models pre-trained only with text data and audio data in our non-parallel corpus respectively. Setting "Late Fusion" concatenates their output representations for downstream fine-tuning. It is worth noting that, the total number of these two unimodal models' parameters equals to that of "Our Method" for comparison purposes. By comparing setting "Late Fusion" with "Our Method", we find our approach still outperforms all three tasks, which proves the importance of introducing inter-modality learning during pre-training phase (the effect of cooperation between IDP and CDAE). Furthermore, by comparing settings "RoBERTa" and "Mockingjay" with settings "w/o Audio Outputs" and "w/o Text Outputs", we can find that using either text outputs or audio outputs from our cross-modal encoders in downstream tasks can achieve better performances than using that from unimodal models, which indicates that each of our cross-modal encoders' outputs already includes information from both modalities, and the best performance is achieved through the fusion of both parts.

% \vspace{-0.2cm}
% \subsection{Data Efficiency in Downstream Tasks}
% We analyze the data efficiency improvement of our pre-training model by visualizing its performance on downstream tasks versus different proportions of training data being used. The results show that our pre-training model only needs half the amount of training data to achieve a better performance than baselines. Figures and more details are provided in Appendix.
\vspace{-0.2cm}
\subsection{Effectiveness Under Semi-supervised Setting}

\vspace{-0.2cm}
\begin{table}[!hbpt]
    \vspace{-0.1cm}
    % \captionsetup{font=small}
    \footnotesize
    \centering
    % \vspace{-0.2cm}
    \scalebox{0.7}{
    \begin{tabular}{@{}ccc|cc|cc|c@{}}
        \toprule
        \multirow{2}{*}{Settings} & \multirow{2}{*}{\begin{tabular}[c]{@{}c@{}}Paired\\ Data\\ (hours)\end{tabular}} & \multirow{2}{*}{\begin{tabular}[c]{@{}c@{}}Unpaired\\ Data\\ (hours)\end{tabular}} & \multicolumn{2}{c|}{\begin{tabular}[c]{@{}c@{}}Emotion\\ Classification\end{tabular}} & \multicolumn{2}{c|}{\begin{tabular}[c]{@{}c@{}}Sentiment\\ Analysis\end{tabular}} & \begin{tabular}[c]{@{}c@{}}Speaker\\ Verification\end{tabular} \\ \cmidrule(l){4-8} 
         &  &  & WA $\uparrow$ & UA $\uparrow$ & MAE $\downarrow$ & Corr $\uparrow$ & EER $\downarrow$ \\ \midrule
        (i) & 360 & 0 & 0.7315 & 0.7362 & 0.6093 & 0.6801 & 0.0192 \\
        (ii) & 360 & 300 & \textbf{0.7352} & \textbf{0.7420} & \textbf{0.6047} & \textbf{0.6807} & \textbf{0.0167} \\ \bottomrule
    \end{tabular}
    }
    \vspace{-0.2cm}
    \caption{The results for demonstrating the effectiveness of our method under semi-supervised setting.}
    \label{tab:rich_resource}
    \vspace{-0.2cm}
\end{table}
% \vspace{-0.2cm}

At last, we further study the effectiveness of our method under semi-supervised setting. We use \textit{train-clean-360} subset in LibriSpeech as our parallel corpus, and form an external non-parallel corpus including 300 hours of speech from remaining training subsets in the same way as mentioned in pre-training details section, the results are presented in Table.~\ref{tab:rich_resource}. From the results, we can see that even if we have sufficient parallel data for pre-training, using our pre-training method can leverage more non-parallel data during pre-training and further boost the performance.

\vspace{-0.4cm}

\section{Conclusion}
% \vspace{-0.2cm}
\label{sec:conclusion}
In this work, we proposed an audio-and-text pre-training framework which leverages only low-resource parallel corpus and extra non-parallel corpus. Our method consists of three main components: intra-modal denoising auto-encoding (IDAE), cross-modal denoising auto-encoding (CDAE) and iterative denoising process (IDP). Our pre-training model achieves comparable performance on multiple downstream tasks compared with the model pre-trained on fully parallel data and outperforms all baselines, demonstrating the great potential of our method. We also show the effectiveness of different components via detailed ablation studies and analysis. For future work, we consider extensions of our method to more paired modalities (i.e., vision-and-language).

\section*{Acknowledgements}
This work was supported in part by National Key R\&D Program of China, under Grant No. 2020AAA0104500 and in part by Beijing Nova Program (Z201100006820068) from Beijing Municipal Science \& Technology Commission.

\bibliographystyle{aaai22}
\bibliography{aaai22}

\end{document}